\documentclass{article}

\usepackage[preprint,dblblindworkshop, final]{neurips_2026}
\workshoptitle{Grounded User Simulation for Model Evaluation and Training:\ Diversity, Fidelity, and Validity}

\usepackage[utf8]{inputenc} 
\usepackage[T1]{fontenc}    
\usepackage{hyperref}       
\usepackage{url}            
\usepackage{booktabs}       
\usepackage{amsfonts}       
\usepackage{nicefrac}       
\usepackage{microtype}      
\usepackage{xcolor}         
\usepackage{graphicx}
\usepackage{subcaption} 
\usepackage{float}
\usepackage[colorinlistoftodos]{todonotes}

\definecolor{mygold}{HTML}{E3C800}
\definecolor{mygoldborder}{HTML}{B09500}

\DeclareRobustCommand*\goldcircled[1]{%
  \tikz[baseline=(char.base)]{
    \node[
      shape=circle,
      draw=mygoldborder ,   
      fill=mygold,   
      text=black,   
      thick,
      inner sep=0.0pt,
      minimum size=10pt, 
    ] (char) {#1};}}

\usepackage{tcolorbox}
\usepackage{xcolor}
\usepackage{tcolorbox}
\usepackage{amsmath}
\tcbuselibrary{skins, breakable}

\definecolor{userblue}{RGB}{235, 243, 250}
\definecolor{agentgray}{RGB}{245, 247, 248}
\definecolor{borderblue}{RGB}{70, 130, 180}
\definecolor{codebg}{RGB}{240, 240, 240}

\title{SWE-DIVE: Multi-Turn Coding Agent Evaluation with Dynamic User Simulation}

\author{%
  Lorenz Brehme \\
  University of Innsbruck\\
  Innsbruck, Austria \\
  \texttt{lorenz.brehme@uibk.ac.at} \\
  \And
  Benjamin Steenhoek \\
  Microsoft \\
  Redmond, WA, USA \\
  \texttt{bensteenhoek@microsoft.com} \\
  \AND
  Ruth Breu \\
   University of Innsbruck\\
  Innsbruck, Austria \\
  \texttt{ruth.breu@uibk.ac.at} \\
}

\begin{document}

\maketitle
\begingroup
\renewcommand{\thefootnote}{}
\footnotetext{Accepted at the NeurIPS 2026 workshop
``Grounded User Simulation for Model Evaluation and Training:
Diversity, Fidelity, and Validity'' (non-archival).}
\endgroup
\begin{abstract}
Coding agents are typically evaluated using one-shot, well-formulated prompts. Benchmarks such as SWE-bench use existing GitHub issues as input for a coding agent and evaluate whether the agent successfully solves the task. However, in real-world use cases, prompts are often not as well-formulated or structured as existing GitHub issues. Additionally, a coding agent session is not limited to a single prompt; it typically involves a multi-turn interaction in which the user iteratively adds new tasks and requirements that the agent must fulfill.

To benchmark these abilities, we developed SWE-DIVE, a dynamic, interactive validation and evaluation framework for benchmarking coding agents in a multi-turn environment. To achieve this, we introduce a user simulator that can simulate different behavioral patterns and communicate with the coding agent. We also introduce a conversation validator that evaluates the user simulator's outputs to determine whether they fulfill the initial requirements and guide the agent toward successfully solving the task. The entire framework is embedded in a Harbor environment, which orchestrates the coding agent, user simulator, and conversation validator, and executes the test environment to determine whether the agent successfully solves the task. Our results and source code are available in our GitHub repository (\url{https://github.com/lorenzbrehme/swe-dive_usersim}).
\end{abstract}

\section{Introduction}

Agents are typically evaluated on benchmarks collected from GitHub issues or crafted for a certain task, like SWE-bench~\citep{jimenez2024swebench}. However, the input format of a complete, detailed github issue is not how users normally interact with agents -- users often give a high-level prompt that is lacking detail and iterate on the prototype until it's complete \citep{baumann2026swechat}. In other words, agent prompting is a process of iteratively defining the problem rather than a one-shot attempt.

\cite{baumann2026swechat} investigated the behavior of humans using a coding agent and found that the agent is used in multi-turn interactions, where different roles emerge, such as the expert nitpicker or the ambiguous programmer. These multi-turn behaviors are already studied in benchmarks such as CodeFlowBench \citep{wang_codeflowbench_2026} and CodeGen \citep{nijkamp2022codegen}, which evaluate coding agents in multi-turn settings. In these benchmarks, tasks typically involve iteratively adding new features, requiring the agent to continually adapt and find solutions. Additionally, AmbigSWE-bench explores the ambiguous behavior of users by providing ambiguous instructions and allowing the agent to ask clarification questions in order to resolve this ambiguity \citep{vijayvargiya2025interactiveagentsovercomeambiguity}.
These multi-turn benchmarks follow a predefined process and do not adapt depending on what the agents ask. \cite{baumann2026swechat} demonstrates that 41\% of the turns were pushbacks, indicating that the agent did not perform as the user expected and that clarification was needed. In addition, \cite{vijayvargiya2025interactiveagentsovercomeambiguity} shows that the agent is still prompted with the full task list, whereas a more realistic setup involves users providing very nitpicky and vague instructions, iteratively adding new tasks in a multi-turn conversation.

To overcome these challenges, we introduce a dynamic interactive validation and evaluation benchmark, SWE-DIVE, that builds upon the DIVE methodology, originally proposed for evaluating retrieval-augmented generation systems \citep{brehme_rag_dive}. SWE-DIVE transforms existing software engineering benchmarks, such as SWE-bench, into dynamic, interactive evaluation environments. It splits the original tasks into incremental sub-tasks and uses an LLM to simulate the user’s discovery process.

\section{Related Work}

Benchmarking software engineering agents requires a collection of tasks consisting of an existing codebase and a corresponding test suite that the agent must satisfy. Benchmarks such as SWE-bench \citep{jimenez2024swebench} and Terminal-Bench \citep{merrill2026terminalbenchbenchmarkingagentshard} provide such tasks by using existing software repositories containing issues that have already been resolved by human developers. Each task includes a predefined initial state, where the provided tests fail, and an expected final state, where the tests pass after the required code modifications have been applied.

These tasks are typically provided to the agent through a single-shot prompt describing the issue to solve. After the agent modifies the codebase, the benchmark executes the associated test suite to determine whether the task has been successfully completed. The resulting success rate is then used as a metric to compare the performance of different software engineering agents.

To extend these benchmarks, approaches such as CodeFlowBench \citep{wang_codeflowbench_2026} and CodeGen \citep{nijkamp2022codegen} introduce multi-turn interactions into the evaluation process. These benchmarks define a sequence of tasks that are provided iteratively to the agent, allowing it to complete multiple development steps over several interactions. Some approaches initialize agents from scratch \citep{nijkamp2022codegen, orlanski_slopcodebench_2025}, while others build upon existing repositories by selecting and filtering issues or introducing additional requirements over multiple iterations, such as in CodeFlowBench \citep{wang_codeflowbench_2026}.
However, these benchmarks remain largely static, as the sequence of tasks and requirements is predefined. They do not capture realistic user behavior, where requirements evolve dynamically, users provide feedback, and goals may change during the development process \citep{baumann2026swechat, vijayvargiya2025interactiveagentsovercomeambiguity}.

Another approach for introducing multi-turn behavior is to provide feedback based on the final outcome of the agent's implementation. In this setting, the agent receives information about whether the tests have passed and, if not, is provided with the test results to iteratively improve its solution \citep{yang2023intercode}. However, these benchmarks still rely on highly structured and formalized task descriptions.

To address this limitation, \cite{garg2025savingswebenchbenchmarkmutation} modifies SWE-bench task descriptions to create more realistic user requests. Similar approaches introduce ambiguity into the task descriptions by incorporating different user personas, where the agent can ask clarification questions to obtain additional information before starting the implementation \citep{vijayvargiya2025interactiveagentsovercomeambiguity, zhou_tom-swe_2026}.

Despite these improvements, an important aspect of realistic software development remains unexplored: users do not only clarify existing requirements but also introduce new tasks and requirements throughout the development process. Existing benchmarks primarily focus on solving a single predefined issue, whereas real-world interactions with coding agents often involve iterative collaboration, where users continuously refine goals and add new requests \citep{baumann2026swechat}. To address this gap, we introduce SWE-DIVE, a new benchmark-agnostic approach for evaluating coding agents through dynamic, multi-turn interactions that better reflect realistic user-agent collaboration.

\section{SWE-DIVE}
\begin{figure}[ht]
  \centering
  \includegraphics[width=\textwidth]{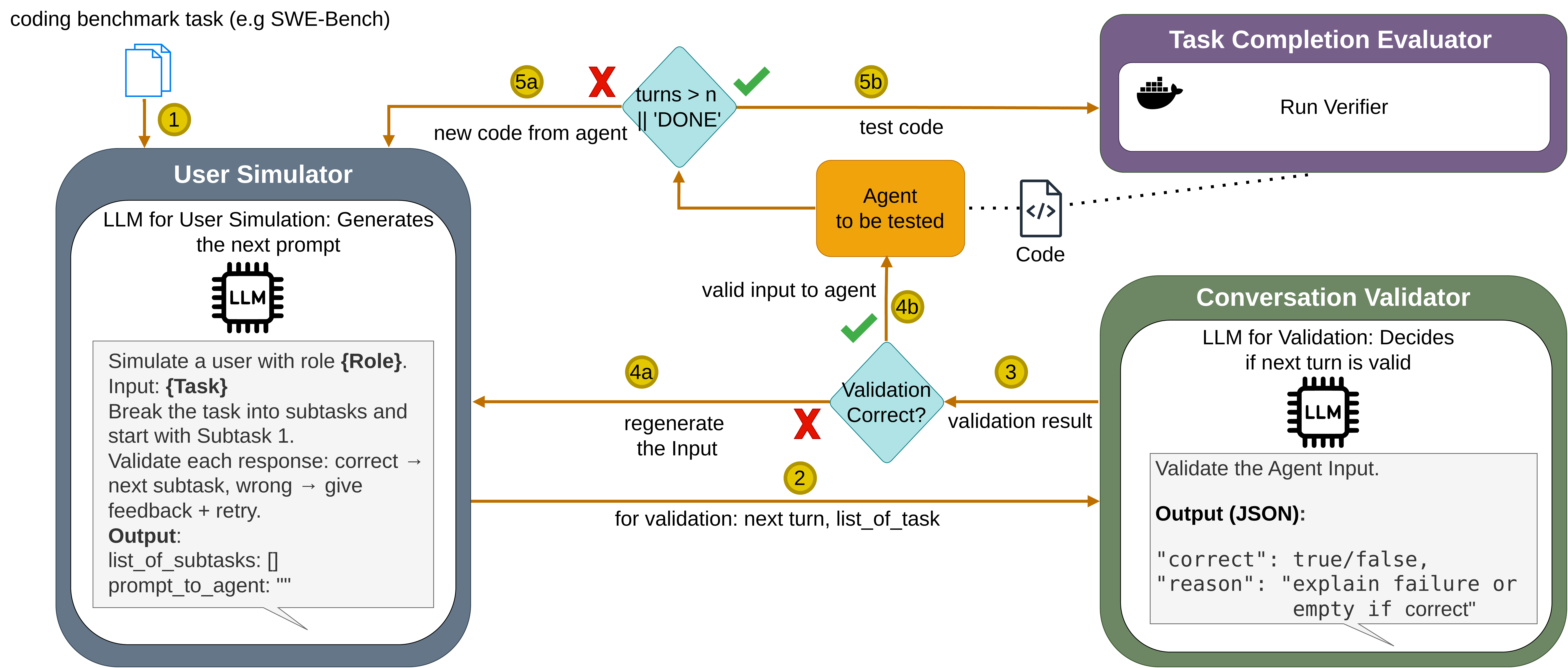}
  \caption{Overview of SWE-DIVE}
\label{fig:triad_appraoch}
\end{figure}

SWE-DIVE is a dynamic, interactive validation and evaluation framework for benchmarking software engineering agents. It builds upon existing benchmarks, such as SWE-bench \citep{jimenez2024swebench} and Terminal-Bench \citep{merrill2026terminalbenchbenchmarkingagentshard}, by transforming single-turn benchmark tasks into multi-turn interactive evaluations. This is achieved through a user simulator that collaborates with the agent to complete the original benchmark tasks.
The framework consists of three main components: the user simulator, the conversation validator, and the Task Completion Evaluator. The user simulator interacts with the evaluated agent, the conversation validator assesses the quality and validity of the interaction, and the Task Completion Evaluator determines whether the benchmark task has been completed successfully.

\subsection{User Simulator}
The first component of SWE-DIVE is the user simulator. This component leverages an LLM to simulate a user interacting with the evaluated software engineering agent. To achieve realistic user interactions, we ground the user simulator in existing, validated task descriptions that specify the underlying requirements and implementation context. This grounding enables the simulator to model a collaborative user who interactively discovers, clarifies, and refines requirements throughout the software implementation process. This distinguishes SWE-DIVE from existing benchmarks \citep{wang_codeflowbench_2026, nijkamp2022codegen}, where interaction is primarily used to provide missing information rather than to model the iterative process of requirement discovery and refinement. The simulator presents the agent with tasks, responds to its questions, provides additional information as needed, and offers feedback based on the agent's responses and modifications to the codebase. It iteratively advances the conversation by introducing follow-up requests or clarifications until the task has been completed successfully or no further progress can be made.

\paragraph{Input: }
Before the interaction begins, the user simulator is assigned a role. We ground our user simulation in behavioral patterns from SWE-Chat \citep{baumann2026swechat}, which were derived from observations of real user interactions with coding agents. Specifically, we leverage three roles. The Expert Nitpicker maintains a stable overall goal while providing a series of precise, targeted corrections to the implementation. Each interaction refines how the agent implements the solution rather than what it is expected to build. The Vague Requester provides broad, underspecified instructions and delegates most implementation decisions to the agent. Finally, the Mind Changer reverses the overall goal during the interaction, fundamentally changing what the agent is expected to build rather than merely modifying the implementation details. These roles are not intended to represent distinct user populations or individual users. Rather, they capture diverse behavioral patterns observed in SWE-Chat and provide controlled conditions for studying how coding agents respond to different user behaviors.

In the first turn, the user simulator is prompted with the initial benchmark task \goldcircled{1}. The complete task specification, such as the issue description from SWE-bench \citep{jimenez2024swebench}, is provided together with the reference solution describing the expected outcome. 
This original objective serves as the grounding for the user simulator, which provides the context needed to guide the interaction. Rather than following a predefined sequence of subtasks, the simulator adaptively determines what information, clarification, or requirement should be communicated next based on the agent's progress. In each turn, the user simulator receives the agent's latest response together with the changes made to the codebase and uses this information to validate and adjust its next message. This enables the simulator to adapt the interaction to the agent's behavior while maintaining alignment with the original task objective.

\paragraph{Output: }
Based on the provided information, the user simulator generates an initial prompt for the agent containing the first subtask. This subtask serves as the starting point and is designed to guide the agent toward completing the overall benchmark task. In addition to the initial prompt, the user simulator generates a complete list of subtasks, providing an overall trajectory for solving the issue.

In subsequent turns, the user simulator generates only the next user prompt. Depending on the agent's progress, this prompt may contain corrections, new subtasks, follow-up requests, or responses to the agent's questions \goldcircled{5a}.

Once the user simulator determines that the overall task has been completed successfully, it returns `done' instead of generating another prompt, thereby terminating the simulation. This allows each agent to complete the task at its own pace \goldcircled{5b}. Additionally, a maximum turn limit is enforced to prevent agents from entering endless interaction loops when they are unable to solve the benchmark task.

\subsection{Conversation Validator}
The conversation validator ensures that the conversation generated by the user simulator remains within the defined trajectory, follows the assigned user role, and effectively guides the agent toward solving the benchmark task. To achieve this, SWE-DIVE implements a validation loop, which can be seen as a modified form of chain-of-verification~\citep{dhuliawala_chain--verification_2023}.

First, the conversation validator receives the output generated by the user simulator \goldcircled{2}. It performs two validation steps: subtask validation and agent input validation. During the initial turn, it validates the generated list of subtasks. This list must cover all necessary steps required to solve the issue, avoid irrelevant tasks, and follow a clear and understandable progression toward the final goal.

In the initial and subsequent turns, the conversation validator evaluates the generated user prompts. Each prompt must remain aligned with the original benchmark issue, follow the defined subtask trajectory, match the assigned user role, and be understandable and actionable for the agent.

For each validation step, the conversation validator returns a boolean value indicating whether the generated output is valid \goldcircled{3}. If an output is considered invalid, the validator additionally provides a reason explaining why the output does not satisfy the requirements and how it can be improved. This feedback is used within the validation loop to prompt the user simulator again, allowing it to regenerate an improved output \goldcircled{4a}. This process continues until the validator approves the generated prompt. Once validated, the prompt is sent to the agent, which executes the requested task and provides its response back to the user simulator for the next interaction cycle \goldcircled{4b}.

\subsection{Task Completion Evaluator}
The Task Completion Evaluator assesses the performance of the software engineering agent and provides quantitative metrics to enable a comprehensive evaluation. It is executed when either the maximum number of turns is reached or the user simulator determines that the conversation is complete. The Task Completion Evaluator leverages the execution environment of the original benchmark to verify whether the assigned task has been successfully completed and calculates the overall success rate.
Upon completion of an evaluation, SWE-DIVE provides a comprehensive performance summary, including the average number of conversation turns, and a detailed analysis of the interactions. Furthermore, the evaluation metrics can be analyzed based on the assigned user role, allowing for a comparison of agent performance under different interaction scenarios. SWE-DIVE also enables filtering based on the number of required turns, providing insights into the efficiency and behavior of agents in realistic, multi-turn software engineering tasks.

\section{Proposed Evaluation Plan}
The framework is implemented using Harbor \citep{Harbor_Framework} and runs in a fully Dockerized environment. As the initial user simulator model, we use \textit{gemini-3.6-flash} \citep{google_gemini-3.6flash_nodate}, while \textit{gemini-2.5-flash-lite} \citep{noauthor_gemini-2.5-flashlite_nodate-1} is used as the validation model. We currently support four coding agents: OpenHands \citep{wang2025openhands}, OpenCode \citep{opencode}, Pi agent \citep{zechner_pi}, and GitHub Copilot \citep{github_copilot}.

Our preliminary evaluation consists of a qualitative analysis of a representative task across all three personas, supplemented by an initial quantitative pilot run on 10 additional tasks per persona to gain early insights. The results and source code are available in our GitHub repository\footnote{\url{https://github.com/lorenzbrehme/swe-dive_usersim}}.

In future work, we plan to evaluate different lightweight open-source models to investigate the performance differences between them and determine whether they are capable of effectively handling this type of interactive coding task. We will also compare the performance of three different user personas: the Expert Nitpicker, the Vague Requester, and the Mind Changer.

Our experiments will use SWE-bench Verified and Terminal-Bench tasks, both of which are supported by the Harbor framework. In general, all benchmarks supported by Harbor are compatible with SWE-DIVE and can be integrated with relatively little effort. Currently, the main limitation is that we use the Git diff as input for the user simulator so that it can assess the agent's progress. This information needs to be available for a benchmark to be directly usable with our framework; otherwise, the Git diff-based approach would need to be replaced with an alternative progress signal provided by the benchmark.

We expect our experiments to yield different performance outcomes due to the personas' varying behaviors. The Expert Nitpicker clearly communicates what they want to accomplish and leaves less room for interpretation, making it easier for the coding agent to identify and implement the required changes. In contrast, the Vague Requester provides less precise instructions, while the Mind Changer introduces changing requirements throughout the interaction.

Beyond the pass/fail status, agent developers need to understand whether a particular interaction style leads to wasteful changes or regressions, such as unnecessary large-scale rewrites. We therefore formalize waste as modifications to files that were not modified in the corresponding golden run and additional tokens spent compared with the golden run. We define regression as tests that fail after the agent's changes but were previously passing.
Based on these definitions, we hypothesize that different user personas will lead to varying levels of waste and regression compared with the baseline SWE-Bench issue.

Additionally, we are interested in comparing the performance of SWE-DIVE against the original benchmark results. Decomposing a task into smaller subtasks through multi-turn interactions could potentially improve the agent's performance and make the benchmark less challenging. This is one of the reasons why we chose to evaluate smaller models rather than state-of-the-art, top-tier models, which already perform very well on the original benchmarks. This allows us to better investigate whether the interactive setting itself provides a performance advantage and how different models respond to the additional complexity introduced by multi-turn interactions.

\subsection{Qualitative Trajectory Analysis}
\label{sec:qualtiy}

\begin{figure}[htbp]
    \centering
    \includegraphics[width=\linewidth]{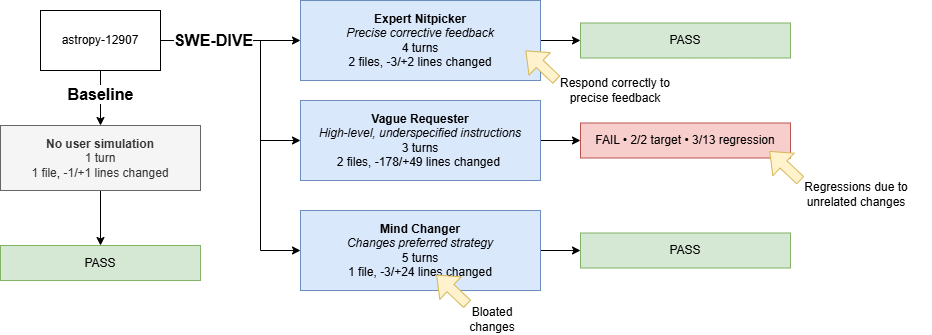}
    \caption{A qualitative case study illustrates the distinct behaviors elicited by each SWE-DIVE persona.}
    \label{fig:case-study-diagram}
\end{figure}

To demonstrate SWE-DIVE in practice, we conduct a qualitative evaluation of agent traces using an SWE-bench Verified \citep{jimenez2024swebench} task (astropy-12907); Figure \ref{fig:case-study-diagram} illustrates the results. The task addresses a bug in the \texttt{separability\_matrix} function, which does not correctly compute separability for nested CompoundModels. We run SWE-DIVE with the \cite{opencode} agent using \textit{gemini-2.5-flash} \citep{google_gemini-2.5flash_nodate} as the underlying model. We evaluate the agent using three personas---Expert Nitpicker, Vague Requester, and Mind Changer---resulting in three user-agent conversations (cf. Figures \ref{fig:qualitative_trace}--\ref{fig:mind_changer_trace}).
We compared with the baseline agent without SWE-DIVE, where the agent resolved the task with a one-file, one-line patch, matching the golden solution.

Each of the three runs exhibited a different behavior that was not exercised by the baseline run.
The agent successfully resolved the issue with both the Expert Nitpicker and Mind Changer personas, whereas the Vague Requester persona led the agent to fix the intended issue but also introduce additional changes that caused 10 previously passing tests to fail.

With the Expert Nitpicker, the agent resolved the issue within four turns. However, the user simulator identified two genuine problems in the agent's implementation: code had been added to the wrong file, and an unrelated leftover change remained in the code. The user simulator provided small, scoped corrections to the existing implementation and then stopped, resulting in a successful resolution. Therefore, this setting tests a behavior that is absent from the single-turn baseline: whether the agent can respond correctly to precise review feedback and recover from implementation detours.
The Vague Requester maintained high-level prompts throughout the interaction, without providing detailed code-level guidance, as intended by the persona. However, the open-ended nature of the requests led the agent to perform a substantial rewrite. Although the intended bug was successfully fixed, the additional changes caused previously passing tests to fail.
Thus, the agent achieved the intended fix but introduced unrelated regressions and changes as a consequence of the persona's vague requests.
Finally, the Mind Changer persona also led the agent to successfully resolve the issue.
In this case, the simulated persona changed its preferred implementation strategy during the interaction, requiring the agent to adapt its approach while continuing to work toward the original objective.

This case study demonstrates how SWE-DIVE tests the agent's behavior under \textit{suboptimal user personas} -- different interaction styles can lead to substantially different agent behavior, and SWE-DIVE can quantify the effects in terms of regressions and wasted tokens.
In this example, the agent appeared to struggle more with the Vague Requester persona than with the more explicitly guiding personas. However, these observations are based on a single task and a small number of runs, and further investigation is required to determine whether this behavior generalizes across tasks and agents.

\subsection{Quantitative Results}

\begin{table}[ht]
\centering

\caption{Summary of results across 11 runs for the three evaluated roles, along with the performance of the original task without a multi-turn conversation. Resolved reports the number of completed tasks; \textit{Passed target tests} reports the number of runs in which all tests transitioned from failing to passing; and \textit{Failed regression test} reports the number in which at least one test transitioned from passing to failing. $\phi$ Turns denotes the average number of agent turns per task. $\Delta$ Tokens shows the change in the mean number of input/output tokens compared to the baseline run. $\Delta$ files shows the change in the mean number of files read/edited compared to the ground-truth patch.}
\begin{tabular}{lrrrr}
\toprule
\textbf{Metric}
& \textbf{Expert Nitpicker}
& \textbf{Mind Changer}
& \textbf{Vague Requester}
& \textbf{Base}\\
\midrule
Resolved
& 4/11 & 4/11 & 3/11 & 3/11\\
Passed target tests
& 6/11 & 6/11 & 4/11 & 4/11 \\
Failed regression test
& 5/11 & 4/11 & 5/11 & 2/11\\
$\phi$ Turns
& 3.27 & 3.36 & 2.73 & 1\\

$\Delta$ Input/Output Tokens
& $+53.5\text{k}/+104.7$
& $+25.5\text{k}/+358.0$
& $+7.1\text{k}/+649.3$ 
& 0/0\\

$\Delta$ Files Read/Edited
& $-0.09/-0.91$
& $+0.36/+0.18$
& $+2.55/+0.18$ 
&0/0\\

\bottomrule
\end{tabular}

\label{tab:summary}
\end{table}

To obtain an initial overview of how the different personas affect agent behavior and performance, we sampled 10 additional tasks and ran SWE-DIVE using the same experimental setup as described in Section \ref{sec:qualtiy}, resulting in a total of 11 evaluated tasks. For each task, we evaluated all three personas and additionally conducted a baseline run in which the agent received the original task description without any multi-turn interaction.

As shown in Table \ref{tab:summary}, all three personas achieved similar success rates, with the Expert Nitpicker and Mind Changer showing slightly better performance than the Vague Requester.

All runs also exhibited regressions, indicating that the agents introduced unnecessary changes that caused previously passing tests to fail. Although the targeted tests transitioned from failing to passing in two cases for the Expert Nitpicker and Mind Changer and in one case for the Vague Requester, the additional changes introduced regressions in the existing test suite. Thus, these cases represent potential successful fixes that were nevertheless accompanied by regressions due to changes beyond the targeted issue. Tables \ref{tab:expert_nitpicker}--\ref{tab:basemode} provide a detailed overview of the failing tests and identify when the agent introduced a regression.

Overall, the Mind Changer exhibited the lowest regression rate, with regressions occurring in 4 of 11 tasks, while both the Expert Nitpicker and Vague Requester resulted in regressions in 5 of 11 tasks. The number of interaction turns also differed slightly between personas. The Mind Changer required the highest number of user requests on average, whereas the Vague Requester required the fewest. However, despite generally requiring fewer turns, the Vague Requester reached the maximum allowed number of turns in one run.

Beyond task success, the personas also enable us to evaluate bloat arising from multi-turn interactions, measured in terms of increased token usage and unrelated changes to the codebase. All personas exhibited increased token usage compared to the baseline, despite submitting multiple turns.\footnote{Note that these averages exhibit high variance, indicating that the additional cost is concentrated in particular tasks rather than occurring uniformly across instances.}
For both Mind Changer and Vague Requester, the agent read and wrote more files compared to the baseline, whereas Expert Nitpicker resulted in decreased file usage. Overall, Vague Requester had both the lowest pass rate and the largest number of unrelated file reads, despite using the fewest turns. It also showed the smallest increase in token usage among the three personas.

\section{Discussion}
Our preliminary results demonstrate how the SWE-DIVE benchmark can be used in practice.
The results reveal that different user personas can surface distinct failure modes and robustness properties that would be missed by evaluating an agent through only single-turn interactions or a single interaction style.

This user simulation enables a diverse range of use cases due to the variety of benchmarks implemented in the Harbor framework \citep{Harbor_Framework}. Additionally, it supports diverse user roles adapted from real-world users and inspired by SWE-Chat \citep{baumann2026swechat}. We define three user roles--Expert Nitpicker, Mind Changer, and Vague Requester--which differ in how users communicate requirements, provide feedback, and evolve their goals throughout the interaction. These roles provide controlled behavioral variation grounded in observed real-user interactions. While this diversity determines the range of user behaviors represented in the benchmark, the fidelity of the simulation depends on how closely these behaviors reflect real user interactions.

SWE-DIVE aims to improve fidelity by modeling users as interactive participants rather than static prompt generators. The user simulator conditions each turn on the agent's latest response and code changes, allowing simulated users to provide corrections, introduce follow-up requirements, or modify their requests as the interaction unfolds. These interactions are grounded in validated benchmark tasks and underlying user behaviors, enabling the user simulator to operate in a multi-turn environment while adopting one of the defined user roles. A direct comparison between SWE-DIVE trajectories and held-out human SWE-Chat trajectories remains an important direction for future work.

Beyond fidelity, the interaction between the simulated user and the agent also raises an important question of validity: whether the use of different simulated users can affect the conclusions drawn from an evaluation. The qualitative trajectories provide an initial indication that this may be the case. On the same Astropy task, different simulated user behaviors resulted in distinct agent trajectories and outcomes: the Vague Requester induced broader changes that introduced regressions, whereas the Expert Nitpicker guided the agent toward a successful resolution through targeted corrections. These results suggest that the simulated user's behavior can influence the observed performance of an agent and, consequently, the conclusions drawn from the evaluation.

 \section{Conclusion}
This work introduces a new benchmark for coding agents that extends the current variety of coding-agent benchmarks to a multi-turn setting with different user personas. It employs a user simulator that communicates with the coding agent in a specific role to guide it toward completing the task. Our preliminary evaluation across 11 tasks demonstrates that SWE-DIVE effectively induces distinct interaction behaviors, uncovering how unconstrained or changing user requirements lead to higher code regression rates.

\section*{Acknowledgements}
The authors used ChatGPT, Grammarly, GitHub Copilot, and Claude Code for language, grammar, and coding support. All content was reviewed and verified by the authors.

\bibliographystyle{apalike}
\bibliography{references.bib}

\appendix
\section{Qualitative Agent Conversations}

\begin{figure}[H]
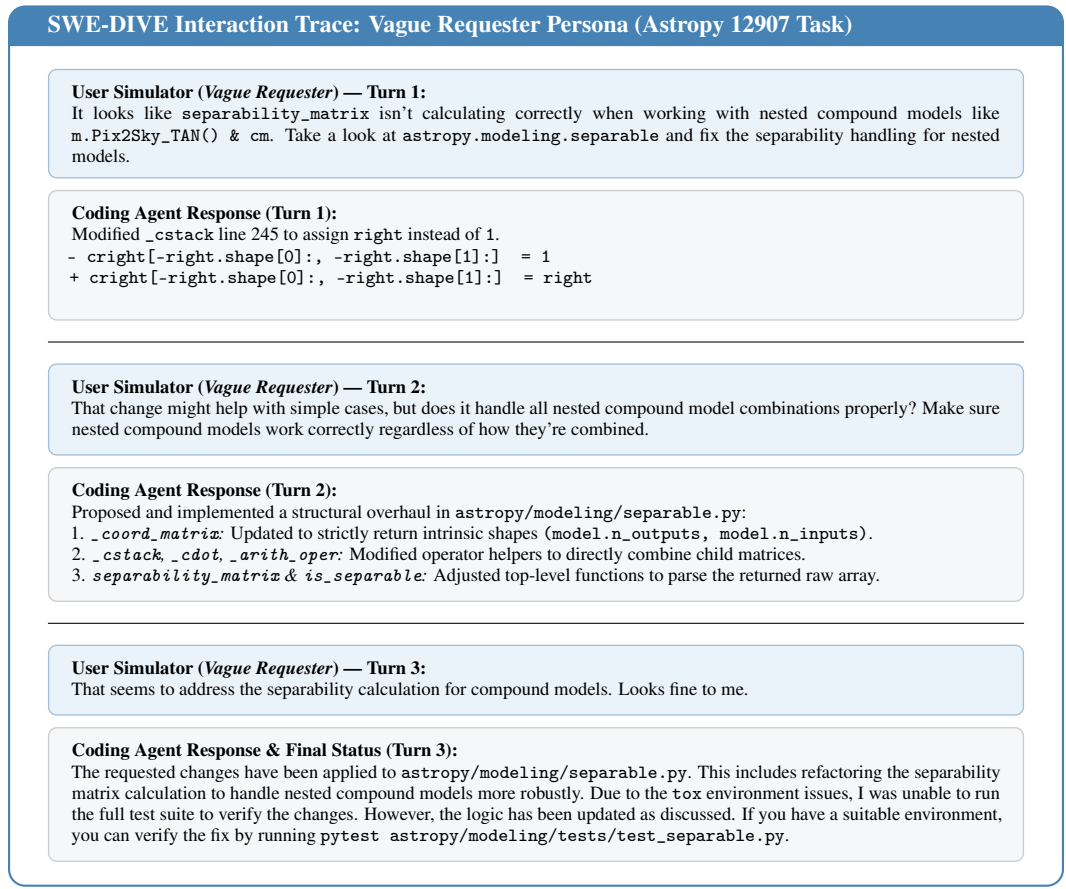

\centering
\begin{tcolorbox}[
    colback=white,
    colframe=borderblue,
    arc=2mm,
    boxrule=0.8pt,
    title=\textbf{SWE-DIVE Interaction Trace: Vague Requester Persona (Astropy 12907 Task)},
    fonttitle=\bfseries\small,
    width=\textwidth
]
\small

\begin{tcolorbox}[colback=userblue, colframe=borderblue!50, boxrule=0.5pt, left=2mm, right=2mm, top=1mm, bottom=1mm, fontupper=\scriptsize]
\textbf{User Simulator (\textit{Vague Requester}) --- Turn 1:} \\
It looks like \texttt{separability\_matrix} isn't calculating correctly when working with nested compound models like \texttt{m.Pix2Sky\_TAN() \& cm}. Take a look at \texttt{astropy.modeling.separable} and fix the separability handling for nested models.
\end{tcolorbox}

\vspace{-2mm}
\begin{tcolorbox}[colback=agentgray, colframe=gray!40, boxrule=0.5pt, left=2mm, right=2mm, top=1mm, bottom=1mm, fontupper=\scriptsize]
\textbf{Coding Agent Response (Turn 1):} \\
Modified \texttt{\_cstack} line 245 to assign \texttt{right} instead of \texttt{1}. \\
\texttt{- cright[-right.shape[0]:, -right.shape[1]:] = 1} \\
\texttt{+ cright[-right.shape[0]:, -right.shape[1]:] = right} \\

\end{tcolorbox}

\vspace{1mm}
\hrule
\vspace{1mm}

\begin{tcolorbox}[colback=userblue, colframe=borderblue!50, boxrule=0.5pt, left=2mm, right=2mm, top=1mm, bottom=1mm, fontupper=\scriptsize]
\textbf{User Simulator (\textit{Vague Requester}) --- Turn 2:} \\
That change might help with simple cases, but does it handle all nested compound model combinations properly? Make sure nested compound models work correctly regardless of how they're combined.
\end{tcolorbox}

\vspace{-2mm}
\begin{tcolorbox}[colback=agentgray, colframe=gray!40, boxrule=0.5pt, left=2mm, right=2mm, top=1mm, bottom=1mm, fontupper=\scriptsize]
\textbf{Coding Agent Response (Turn 2):} \\
Proposed and implemented a structural overhaul in \texttt{astropy/modeling/separable.py}: \\
1. \textit{\texttt{\_coord\_matrix}:} Updated to strictly return intrinsic shapes \texttt{(model.n\_outputs, model.n\_inputs)}. \\
2. \textit{\texttt{\_cstack}, \texttt{\_cdot}, \texttt{\_arith\_oper}:} Modified operator helpers to directly combine child matrices. \\
3. \textit{\texttt{separability\_matrix} \& \texttt{is\_separable}:} Adjusted top-level functions to parse the returned raw array.
\end{tcolorbox}

\vspace{1mm}
\hrule
\vspace{1mm}

\begin{tcolorbox}[colback=userblue, colframe=borderblue!50, boxrule=0.5pt, left=2mm, right=2mm, top=1mm, bottom=1mm, fontupper=\scriptsize]
\textbf{User Simulator (\textit{Vague Requester}) --- Turn 3:} \\
That seems to address the separability calculation for compound models. Looks fine to me.
\end{tcolorbox}

\vspace{-2mm}
\begin{tcolorbox}[colback=agentgray, colframe=gray!40, boxrule=0.5pt, left=2mm, right=2mm, top=1mm, bottom=1mm, fontupper=\scriptsize]
\textbf{Coding Agent Response \& Final Status (Turn 3):} \\
The requested changes have been applied to \texttt{astropy/modeling/separable.py}. This includes refactoring the separability matrix calculation to handle nested compound models more robustly. Due to the \texttt{tox} environment issues, I was unable to run the full test suite to verify the changes. However, the logic has been updated as discussed. If you have a suitable environment, you can verify the fix by running \texttt{pytest astropy/modeling/tests/test\_separable.py}.
\end{tcolorbox}

\end{tcolorbox}
\caption{Representative multi-turn trajectory in SWE-DIVE with a fixed Vague Requester persona}
\label{fig:qualitative_trace}
\end{figure}

\begin{figure}[H]
\centering
\begin{tcolorbox}[
    colback=white,
    colframe=borderblue,
    arc=2mm,
    boxrule=0.8pt,
    title=\textbf{SWE-DIVE Interaction Trace: Expert Nitpicker Persona (Astropy 12907 Task)},
    fonttitle=\bfseries\small,
    width=\textwidth
]
\small

\begin{tcolorbox}[colback=userblue, colframe=borderblue!50, boxrule=0.5pt, left=2mm, right=2mm, top=1mm, bottom=1mm, fontupper=\scriptsize]
\textbf{User Simulator (\textit{Expert Nitpicker}) --- Turn 1:} \\
Take a look at \texttt{separability\_matrix} in \texttt{astropy/modeling/separable.py}. When dealing with a nested \texttt{CompoundModel} like \texttt{m.Pix2Sky\_TAN() \& cm}, \texttt{\_separable} returns fully coupled matrices for nested submodels instead of preserving their internal separability. Start by inspecting \texttt{\_separable} and how \texttt{\_compound} handles nested trees.
\end{tcolorbox}

\vspace{-2mm}
\begin{tcolorbox}[colback=agentgray, colframe=gray!40, boxrule=0.5pt, left=2mm, right=2mm, top=1mm, bottom=1mm, fontupper=\scriptsize]
\textbf{Coding Agent Response (Turn 1):} \\
Added \texttt{\_calculate\_separability\_matrix} to \texttt{CompoundModel} in \texttt{astropy/modeling/core.py} to recurse into submodels. \\
Attempted running tests, but encountered environmental compilation errors (\texttt{longintrepr.h}).
\end{tcolorbox}

\vspace{1mm}
\hrule
\vspace{1mm}

\begin{tcolorbox}[colback=userblue, colframe=borderblue!50, boxrule=0.5pt, left=2mm, right=2mm, top=1mm, bottom=1mm, fontupper=\scriptsize]
\textbf{User Simulator (\textit{Expert Nitpicker}) --- Turn 2:} \\
Don't add \texttt{\_calculate\_separability\_matrix} to \texttt{CompoundModel} in \texttt{core.py} --- keep all separability logic inside \texttt{astropy/modeling/separable.py}. Also revert the changes to \texttt{tox.ini}.
\end{tcolorbox}

\vspace{-2mm}
\begin{tcolorbox}[colback=agentgray, colframe=gray!40, boxrule=0.5pt, left=2mm, right=2mm, top=1mm, bottom=1mm, fontupper=\scriptsize]
\textbf{Coding Agent Response (Turn 2):} \\
Reverted \texttt{core.py} and \texttt{tox.ini} changes. Instead, updated line 245 of \texttt{\_cstack} in \texttt{astropy/modeling/separable.py}: \\
\texttt{- cright[-right.shape[0]:, -right.shape[1]:] = 1} \\
\texttt{+ cright[-right.shape[0]:, -right.shape[1]:] = right}
\end{tcolorbox}

\vspace{1mm}
\hrule
\vspace{1mm}

\begin{tcolorbox}[colback=userblue, colframe=borderblue!50, boxrule=0.5pt, left=2mm, right=2mm, top=1mm, bottom=1mm, fontupper=\scriptsize]
\textbf{User Simulator (\textit{Expert Nitpicker}) --- Turn 3:} \\
\texttt{tox.ini} is still modified in your diff, revert that change completely. Also in \texttt{\_cstack}, don't slice with \texttt{-right.shape[0]:} when assigning \texttt{right}, just assign \texttt{cright[-right.shape[0]:, :] = right} directly.
\end{tcolorbox}

\vspace{-2mm}
\begin{tcolorbox}[colback=agentgray, colframe=gray!40, boxrule=0.5pt, left=2mm, right=2mm, top=1mm, bottom=1mm, fontupper=\scriptsize]
\textbf{Coding Agent Response (Turn 3):} \\
 Reverted \texttt{tox.ini} formatting back to space-delimited \texttt{passenv}. Updated \texttt{\_cstack} slicing to \texttt{cright[-right.shape[0]:, :] = right} as instructed.
\end{tcolorbox}

\vspace{1mm}
\hrule
\vspace{1mm}

\begin{tcolorbox}[colback=userblue, colframe=borderblue!50, boxrule=0.5pt, left=2mm, right=2mm, top=1mm, bottom=1mm, fontupper=\scriptsize]
\textbf{User Simulator (\textit{Expert Nitpicker}) --- Turn 4:} \\
In \texttt{\_cstack}, also update \texttt{cleft} handling so it correctly uses \texttt{left} matrix instead of initializing with ones when \texttt{isinstance(left, np.ndarray)} is false.
\end{tcolorbox}

\vspace{-2mm}
\begin{tcolorbox}[colback=agentgray, colframe=gray!40, boxrule=0.5pt, left=2mm, right=2mm, top=1mm, bottom=1mm, fontupper=\scriptsize]
\textbf{Coding Agent Response \& Final Diff (Turn 4):} \\
Updated \texttt{cleft[: left.shape[0], :] = left} in \texttt{\_cstack} for consistency with \texttt{cright}. \\
 Cleaned up code edits in \texttt{astropy/modeling/separable.py}.
\end{tcolorbox}

\end{tcolorbox}
\caption{Representative multi-turn trajectory in SWE-DIVE with a fixed Expert Nitpicker persona.}
\label{fig:expert_nitpicker}
\end{figure}

\begin{figure}[H]
\centering
\begin{tcolorbox}[
    colback=white,
    colframe=borderblue,
    arc=2mm,
    boxrule=0.8pt,
    title=\textbf{SWE-DIVE Interaction Trace: Mind Changer Persona (Astropy 12907 Task)},
    fonttitle=\bfseries\small,
    width=\textwidth
]
\small
\begin{tcolorbox}[colback=userblue, colframe=borderblue!50, boxrule=0.5pt, left=2mm, right=2mm, top=1mm, bottom=1mm, fontupper=\scriptsize]
\textbf{User Simulator (\textit{Mind Changer}) --- Turn 1:} \\
Take a look at \texttt{astropy/modeling/separable.py} where \texttt{separability\_matrix} fails on nested compound models like \texttt{m.Pix2Sky\_TAN() \& (m.Linear1D(10) \& m.Linear1D(5))}. Start by updating \texttt{\_separable} to recursively process child models when a \texttt{CompoundModel} contains another \texttt{CompoundModel}.
\end{tcolorbox}

\vspace{-2mm}
\begin{tcolorbox}[colback=agentgray, colframe=gray!40, boxrule=0.5pt, left=2mm, right=2mm, top=1mm, bottom=1mm, fontupper=\scriptsize]
\textbf{Coding Agent Response (Turn 1):} \\
I've updated \texttt{astropy/modeling/separable.py}. The \texttt{\_cstack} function now correctly handles nested compound models by preserving the separability matrix from the right-hand side of the \texttt{\&} operation. This should resolve the issue with \texttt{separability\_matrix} failing on nested compound models.
\end{tcolorbox}

\vspace{1mm}
\hrule
\vspace{1mm}

\begin{tcolorbox}[colback=userblue, colframe=borderblue!50, boxrule=0.5pt, left=2mm, right=2mm, top=1mm, bottom=1mm, fontupper=\scriptsize]
\textbf{User Simulator (\textit{Mind Changer}) --- Turn 2} \\
Wait, looking at \texttt{\_cstack} in \texttt{astropy/modeling/separable.py}, patching \texttt{cright} directly like that is missing the point. Actually, thinking about this again: we shouldn't be trying to recurse or patch matrix combination logic in \texttt{\_separable}. Instead, let's completely pivot: modify \texttt{\_compound} to flatten the nested compound model operators tree first before building any separability matrices.
\end{tcolorbox}

\vspace{-2mm}
\begin{tcolorbox}[colback=agentgray, colframe=gray!40, boxrule=0.5pt, left=2mm, right=2mm, top=1mm, bottom=1mm, fontupper=\scriptsize]
\textbf{Coding Agent Response (Turn 2):} \\
I've updated \texttt{astropy/modeling/separable.py}. The \texttt{\_separable} function now flattens nested \texttt{CompoundModel}`s with the same operator before computing the separability matrix. This should correctly handle cases like \texttt{m.Pix2Sky\_TAN() \& (m.Linear1D(10) \& m.Linear1D(5))}.",
   
\end{tcolorbox}

\vspace{1mm}
\hrule
\vspace{1mm}

\begin{tcolorbox}[colback=userblue, colframe=borderblue!50, boxrule=0.5pt, left=2mm, right=2mm, top=1mm, bottom=1mm, fontupper=\scriptsize]
\textbf{User Simulator (\textit{Mind Changer}) --- Turn 3:} \\
You put the loop in \texttt{\_separable}, but \texttt{\_compound} is where complex compound trees are evaluated. Move the flattening handling into \texttt{\_compound} so that \texttt{\_separable} simply calls \texttt{\_compound} on compound models, and make sure \texttt{\_cstack} and \texttt{\_hstack} are called properly with correct matrix dimension offsets.
\end{tcolorbox}

\vspace{-2mm}
\begin{tcolorbox}[colback=agentgray, colframe=gray!40, boxrule=0.5pt, left=2mm, right=2mm, top=1mm, bottom=1mm, fontupper=\scriptsize]
\textbf{Coding Agent Response (Turn 3):} \\
I've moved the flattening logic for compound models into a new \texttt{\_compound} function, and \texttt{\_separable} now calls \texttt{\_compound} for \texttt{CompoundModel} instances. This ensures that nested compound models are flattened before their separability matrices are combined. The existing \texttt{\_cstack} and \texttt{\_cdot} functions already handle matrix dimension offsets correctly when combining the results from \texttt{\_separable} on sub-models.
\end{tcolorbox}

\vspace{1mm}
\hrule
\vspace{1mm}

\begin{tcolorbox}[colback=userblue, colframe=borderblue!50, boxrule=0.5pt, left=2mm, right=2mm, top=1mm, bottom=1mm, fontupper=\scriptsize]
\textbf{User Simulator (\textit{Mind Changer}) --- Turn 4:} \\
In \texttt{\_compound}, you left duplicate \texttt{elif isinstance(transform, Model):} cleanup code hanging at the bottom after \texttt{return result\_matrix}. Clean up that unreachable trailing code block.
\end{tcolorbox}

\vspace{-2mm}
\begin{tcolorbox}[colback=agentgray, colframe=gray!40, boxrule=0.5pt, left=2mm, right=2mm, top=1mm, bottom=1mm, fontupper=\scriptsize]
\textbf{Coding Agent Response (Turn 4):} \\
Reviewed \texttt{\_compound} definition and questioned where the dangling block was located, confirming that \texttt{\_compound} ends cleanly with \texttt{return result\_matrix}. Could you please clarify where exactly you are seeing this duplicate or unreachable code?
\end{tcolorbox}

\vspace{1mm}
\hrule
\vspace{1mm}

\begin{tcolorbox}[colback=userblue, colframe=borderblue!50, boxrule=0.5pt, left=2mm, right=2mm, top=1mm, bottom=1mm, fontupper=\scriptsize]
\textbf{User Simulator (\textit{Mind Changer}) --- Turn 5:} \\
Look at lines 332-333 in the diff right below \texttt{return result\_matrix} --- there is a leftover \texttt{elif isinstance(transform, Model):} dangling at module level outside \texttt{\_compound}. Delete those lines.
\end{tcolorbox}

\vspace{-2mm}
\begin{tcolorbox}[colback=agentgray, colframe=gray!40, boxrule=0.5pt, left=2mm, right=2mm, top=1mm, bottom=1mm, fontupper=\scriptsize]
\textbf{Coding Agent Response \& Cleanup (Turn 5):} \\
You are absolutely correct! My apologies for missing that dangling code block. I've removed the unreachable \texttt{elif isinstance(transform, Model):} block from \texttt{astropy/modeling/separable.py}. Thank you for catching that!
\end{tcolorbox}
\end{tcolorbox}
\caption{Representative multi-turn trajectory in SWE-DIVE with a fixed Mind Changer persona.}
\label{fig:mind_changer_trace}
\end{figure}

\newpage
\section{Detailed Results}
\begin{table}[ht]
\centering
\caption{Results across 11 runs with the Expert Nitpicker. Resolved indicates whether the task was successfully completed. Fail-to-pass reports the number of tests that transitioned from failing to passing as a result of the agent’s changes, while pass-to-fail reports the number of previously passing tests that became failing. Turns denotes the number of agent turns required to complete the task, and Task identifies the task evaluated.}
\begin{tabular}{lllll}
\toprule
\textbf{Task} & \textbf{Resolved} & \textbf{FAIL\_TO\_PASS} & \textbf{PASS\_TO\_FAIL} & \textbf{Turns} \\
\midrule
astropy-12907    & true  & 2/2  & 0/13       & 4 \\
matplotlib-22865 & true  & 3/3  & 0/57       & 3 \\
sympy-16886      & true  & 1/1  & 0/42       & 2 \\
requests-1142    & true  & 1/1  & 0/5       & 2 \\
django-16938     & false & 18/23 & 0/65      & 2 \\
django-10554     & false & 0/2  & 1/23       & 6 \\
django-14311     & false & 1/1  & 3/59       & 2 \\
django-16032     & false & 0/2  & 1/77       & 3 \\
django-16950     & false & 1/1  & 3/3 & 3 \\
matplotlib-14623 & false & 0/1  & 0/400       & 4 \\
pytest-7236      & false & 0/1  & 1/51       & 5 \\
\midrule
\textbf{Total} & 4/11 & 6/11 & 5/11 & $\phi$ 3.27 \\
\bottomrule
\end{tabular}
\label{tab:expert_nitpicker}
\end{table}

\begin{table}[ht]
\centering
\caption{Results across 11 runs with the Mind Changer. Resolved indicates whether the task was successfully completed. Fail-to-pass reports the number of tests that transitioned from failing to passing as a result of the agent’s changes, while pass-to-fail reports the number of previously passing tests that became failing. Turns denotes the number of agent turns required to complete the task, and Task identifies the task evaluated.}
\begin{tabular}{lllll}
\toprule
\textbf{Task} & \textbf{Resolved} & \textbf{FAIL\_TO\_PASS} & \textbf{PASS\_TO\_FAIL} & \textbf{Turns} \\
\midrule
astropy-12907    & true  & 2/2  & 0/13       & 5 \\
matplotlib-22865 & false & 1/3   & 1/57  & 2 \\
sympy-16886      & true  & 1/1   & 0/42  & 2 \\
requests-1142    & true  & 1/1   & 0/5  & 4\\
django-16938     & true  & 23/23 & 0/65  & 3 \\
django-10554     & false & 0/2   & 1/23  & 3 \\
django-14311     & false & 1/1   & 3/59  & 3 \\
django-16032     & false & 2/2   & 17/77 & 5 \\
django-16950     & false & 0/1   & 0/3  & 4 \\
matplotlib-14623 & false & 0/1   & 0/400  & 3 \\
pytest-7236      & false & 0/1   & 0/51  & 3 \\
\midrule
\textbf{Total} & 4/11 & 6/11 & 4/11 & $\phi$ 3.36 \\
\bottomrule
\end{tabular}
\label{tab:mindchanger}
\end{table}

\begin{table}[ht]
\centering
\caption{Results across 11 runs with the Vague Requester. Resolved indicates whether the task was successfully completed. Fail-to-pass reports the number of tests that transitioned from failing to passing as a result of the agent’s changes, while pass-to-fail reports the number of previously passing tests that became failing. Turns denotes the number of agent turns required to complete the task, and Task identifies the task evaluated.}
\begin{tabular}{lllll}
\toprule
\textbf{Task} & \textbf{Resolved} & \textbf{FAIL\_TO\_PASS} & \textbf{PASS\_TO\_FAIL} & \textbf{Turns} \\
\midrule
astropy-12907    & false  & 2/2  &  10/13     & 3 \\
matplotlib-22865 & false & 1/3   & 1/57 & 3 \\
sympy-16886      & true  & 1/1   & 0/42 & 1 \\
requests-1142    & true  & 1/1   & 0/5 & 3 \\
django-16938     & true  & 23/23 & 0/65 & 1 \\
django-10554     & false & 0/2   & 0/23 & 1 \\
django-14311     & false & 0/1   & 3/59 & 2 \\
django-16032     & false & 0/2   & 0/77 & 1 \\
django-16950     & false & 0/1   & 2/3 & 3 \\
matplotlib-14623 & false & 0/1   & 0/400 & 10 (max) \\
pytest-7236      & false & 0/1   & 1/51 & 2 \\
\midrule
\textbf{Total} & 3/11 & 4/11 & 5/11 & $\phi$ 2.73 \\
\bottomrule
\end{tabular}
\label{tab:vaguerequester}
\end{table}

\begin{table}[ht]
\centering
\caption{Results across 11 runs using the base task without multi-turn behavior. Resolved indicates whether the task was successfully completed. Fail-to-pass reports the number of tests that transitioned from failing to passing as a result of the agent’s changes, while pass-to-fail reports the number of previously passing tests that became failing. Turns denotes the number of agent turns required to complete the task, and Task identifies the task evaluated.}
\begin{tabular}{lllll}
\toprule
\textbf{Task} & \textbf{Resolved} & \textbf{FAIL\_TO\_PASS} & \textbf{PASS\_TO\_FAIL} & \textbf{Turns} \\
\midrule
astropy-12907    & true  & 2/2  &  0/13     & 1 \\
matplotlib-22865 & false & 1/3   & 1/57 & 1 \\
sympy-16886      & true  & 1/1   & 0/42 & 1 \\
requests-1142    & false & 0/1   & 0/5 & 1 \\
django-16938     & false  & 0/23 & 0/65 & 1 \\
django-10554     & false & 0/2   & 0/23 & 1 \\
django-14311     & false & 1/1   & 3/59 & 1 \\
django-16032     & true & 2/2   & 0/77 & 1 \\
django-16950     & false & 0/1   & 0/3 & 1 \\
matplotlib-14623 & false & 0/1   & 0/400 & 1 \\
pytest-7236      & false & 0/1   & 0/51 & 1 \\
\midrule
\textbf{Total} & 3/11 & 4/11 & 2/11 & $\phi$ 1 \\
\bottomrule
\end{tabular}
\label{tab:basemode}
\end{table}

\end{document}